\documentclass[a4paper,fleqn,usenatbib,letters]{mnras}

\usepackage{newtxtext,newtxmath}

\usepackage[T1]{fontenc}

\usepackage{graphicx}	
\usepackage{amsmath}	
\usepackage{longtable}
\usepackage{siunitx}
\DeclareSIUnit\year{yr}
\DeclareSIUnit\mag{mag}
\usepackage{tablefootnote}
\usepackage{xspace}
\usepackage{booktabs}
\usepackage{xcolor}
\usepackage[normalem]{ulem}
\usepackage{hhline}
\usepackage{hyperref}

\numberwithin{equation}{section}
\graphicspath{{./}{figures/}}
\defcitealias{Rose2019}{R19}
\defcitealias{Rose2020}{R20}
\defcitealias{Lee2020}{L20}
\defcitealias{Kang2020}{K20}
\defcitealias{Uddin_2020}{U20}
\defcitealias{Z20_letter}{Z21L}
\newcommand{\rev}[1]{\textcolor{black}{#1}}
\newcommand{\RoseMC}{\citetalias{Rose2019}\xspace}
\newcommand{\RoseSN}{\citetalias{Rose2020}\xspace}
\newcommand{\Lee}{\citetalias{Lee2020}\xspace}
\newcommand{\Kang}{\citetalias{Kang2020}\xspace}
\newcommand{\Uddin}{\citetalias{Uddin_2020}\xspace}
\newcommand{\ZL}{\citetalias{Z20_letter}\xspace}
\newcommand{\LCDM}{$\Lambda$CDM\xspace}
\newcommand{\Nsamples}{{$\sim200$}\xspace}

\title[SN~Ia luminosity and host-galaxy properties]{On the relationship between Type Ia supernova luminosity and host-galaxy properties}

\author[Y. S. Murakami et al.]{Yukei S. Murakami,$^{1,2,3}$  \thanks{E-mail: sterling.astro@berkeley.edu}
Benjamin E. Stahl,$^{1,2,4}$
Keto D. Zhang,$^{1}$
\newauthor
Matthew R. Chu$^{2}$,
Emma C. McGinness$^{2}$,
Kishore C. Patra$^{1,5}$,
and Alexei V. Filippenko$^{1,6,7}$
\\
$^{1}$Department of Astronomy, University of California, Berkeley, CA 94720-3411, USA\\
$^{2}$Department of Physics, University of California, Berkeley, CA 94720-7300, USA\\
$^{3}$Google Lick Predoctoral Fellow\\
$^{4}$Marc J. Staley Graduate Fellow\\
$^{5}$Nagaraj-Noll Graduate Fellow \\
$^{6}$Miller Institute for Basic Research in Science, University of California, Berkeley, CA 94720, USA\\
$^{7}$Miller Senior Fellow
}

\date{Accepted to MNRAS.}

\pubyear{2020}

\begin{document}

\label{firstpage}
\pagerange{\pageref{firstpage}--\pageref{lastpage}}
\maketitle

\begin{abstract}
    A string of recent studies has debated the \rev{exact form and physical origin} of an evolutionary trend between the peak luminosity of 
    Type Ia supernovae (SNe~Ia) and the properties of the galaxies that host them.
    We shed new light on the discussion by presenting an analysis of \Nsamples low-redshift SNe~Ia in which we measure the separation of Hubble residuals (HR; as probes of luminosity) between two host-galaxy morphological types. 
    We show that this separation can test the predictions made by recently proposed models, using an independently and empirically determined distribution of each morphological type in host-property space.
    Our results are partially consistent with the new HR--age slope,
    \rev{but we find significant scatter in the predictions from different galaxy catalogues.
    The inconsistency in age illuminates an issue in the current debate that was not obvious in the long-discussed mass models: HR--host-property models are strongly dependent on the methods employed to determine galaxy properties.}
    \rev{While our results demonstrate the difficulty in constructing a universal model for age as a proxy for host environment, our results indeed identify evolutionary trends between mass, age, morphology, and HR values, encouraging (or requiring, if such trends are to be accounted for in cosmological studies) further investigation}. 
\end{abstract}

\begin{keywords}
    distance scale -- cosmology: observations -- supernovae: general -- methods: data analysis -- methods: statistical
\end{keywords}


\section{Introduction} 
\label{sec: introduction}
    Type Ia supernovae (SNe~Ia), brilliant explosions of white dwarfs 
    \citep[see, e.g.,][for reviews]{Maoz_2014_SNeprojReview,Jha_2019_SNeIareview}, produce a remarkably small range of peak luminosities. Moreover, this range can be further narrowed through empirical relationships that quantify the correlated rate of photometric evolution \citep[e.g.,][]{Phillips1993,Riess1996,mlcs2k2}. Consequently, SNe~Ia are excellent \emph{standardisable} candles for measuring extragalactic distances. 
    Indeed, evidence for the accelerating expansion of the Universe was first found by studying distances derived in this way \citep{Riess1998,Perlmutter1999} --- a paradigm-shifting discovery that has been upheld and verified by independent observations at different redshift ranges \citep[e.g.,][]{Planck2018_2020A&A}. Today, SNe~Ia remain one of the best diagnostic tools for testing \LCDM cosmology \citep[for reviews, see, e.g.,][]{Filippenko_2005_review,Riess_2019_Nature}.
    
\begin{figure}
        \centering
        \includegraphics[width=\linewidth]{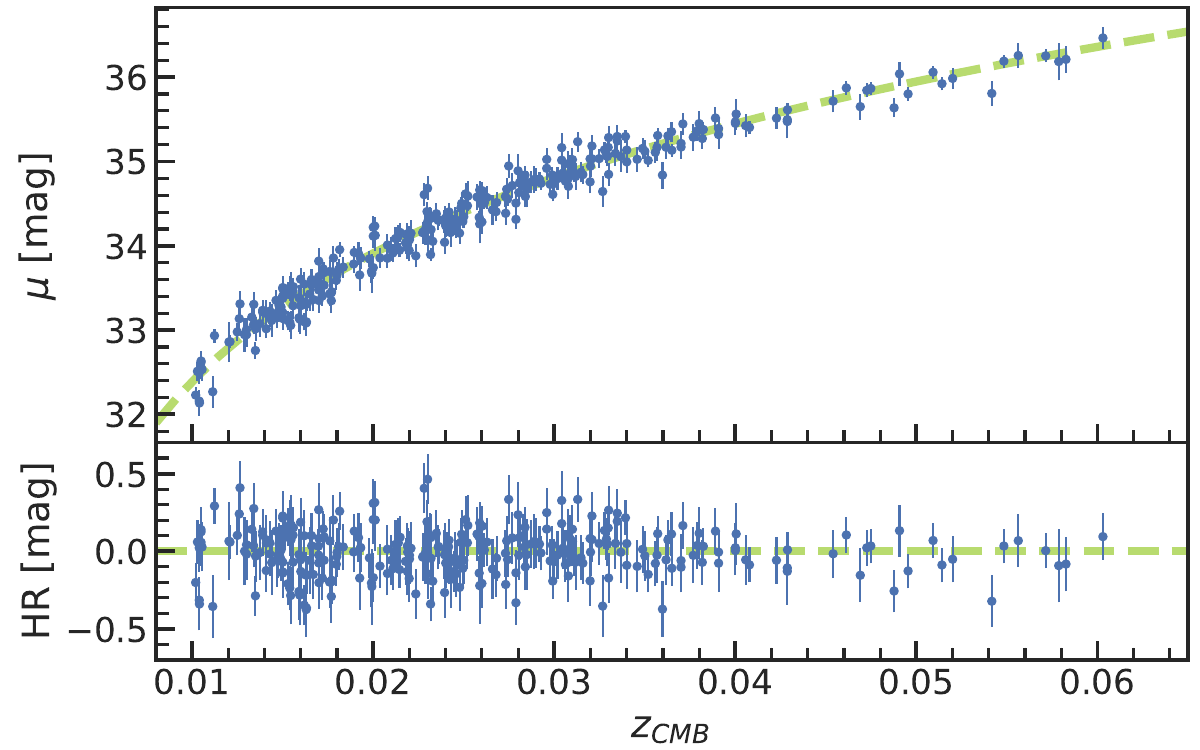}
        \caption{Hubble diagram of our SN~Ia sample (upper panel), with HR (lower panel) derived using fiducial cosmology as prescribed in Sec.~\ref{sec:data-SNe}.
        }
        \label{fig:HR_cosmo}
    \end{figure}

    In roughly the past decade, correlations have been discovered between the luminosity and host-galaxy environments of SNe~Ia \citep[e.g.,][]{Childress_2013_Mstep}. Left unresolved, they could manifest as biases in the cosmological statements that follow from SN~Ia observations; thus, understanding (and if necessary, correcting) them is essential.
    As a notable example, \cite{Childress_2013_Mstep} suggest that an offset in Hubble residual\footnote{The deviation of each SN's distance modulus from the cosmological value at its redshift. For details, see Sec.~\ref{sec:data-SNe} and the lower panel of Figure~\ref{fig:HR_cosmo}.} (HR) values
    should be introduced to account for the differences that manifest --- in aggregate --- between SNe~Ia in ``light'' and ``heavy'' galaxies. This ``mass step'' at $\sim10^{10}\, M_\odot$, now commonplace in cosmological studies \citep[e.g.,][]{Betoule2014,Scolnic2018}, makes SNe~Ia in ``heavier'' galaxies appear brighter by $\sim 0.05$--0.1\,mag. Alternatively, a more continuous model (``mass slope'') has also been proposed \citep[e.g.,][hereafter \Uddin]{Uddin_2020} in the search for a more physical form of the HR--mass relation.

    More recently, a study of SNe~Ia with early-type host galaxies has proposed a large and significant correlation between the (spectroscopically determined) galaxy-luminosity-weighted age 
    and HR \citep[][hereafter \Kang]{Kang2020}. That study, which is based on $\sim30$ SNe from early-type galaxies (E--S0), finds a linear relationship between age and HR with a slope that (if correct) predicts a large luminosity offset between SN~Ia populations from young and old host galaxies. In response, another study \citep[][hereafter \RoseSN]{Rose2020} points out that this result disagrees with their prior analysis that uses photometrically derived host-galaxy ages \citep[][hereafter \RoseMC]{Rose2019} and finds a significantly smaller (if not zero) slope. \RoseSN also suggest that a similar conclusion can be drawn from \cite{Jones_2018_SNenv}.
    Responding to \RoseSN, \citet[][hereafter \Lee]{Lee2020} continue to disagree and claim a large slope comparable to their original result from \Kang using the data from \RoseMC.
    In our separate work, \citet[][hereafter \ZL]{Z20_letter} demonstrated that the statistical analysis used by \Lee in deriving a slope is considerably flawed. 
    
    In this {\it Letter}, we provide an independent test of some of the aforementioned relations between HR and host-galaxy properties (e.g., mass and age) \rev{and their \emph{universality}}. 
    As a proxy for direct measurements of host-galaxy \rev{properties}, we use Hubble morphology types which are known to be correlated with both the ages and the masses of the galaxies that they classify \citep[for reviews, see, e.g.,][]{Roberts_1994_Morph_vs_params,Conselice_2014_galaxy_review}. 
    By dividing a sample of cosmologically viable SNe~Ia between two galaxy population groups --- early-type (old galaxies) and late-type (young galaxies) --- we evaluate the difference between their corresponding HR distributions. Using the separation between two galaxy types in HR distributions as a probe, we test the recently reported HR--age slope by \Kang \rev{and \ZL} and the HR--mass slope by \cite{Uddin_2020}.
    This analysis takes advantage of spectroscopically \rev{and photometrically} determined host-galaxy properties derived via multiple data sets and methodologies to construct the HR distribution for SNe~Ia predicted by such slopes.
    We discuss our SN and galaxy dataset in Sections~\ref{sec: Data}~and~\ref{sec:empirical_prediction}, and provide the results and conclusions in later sections.

\section{Data}
\label{sec: Data}

    \begin{figure*}
        \centering
        \includegraphics[width=\linewidth]{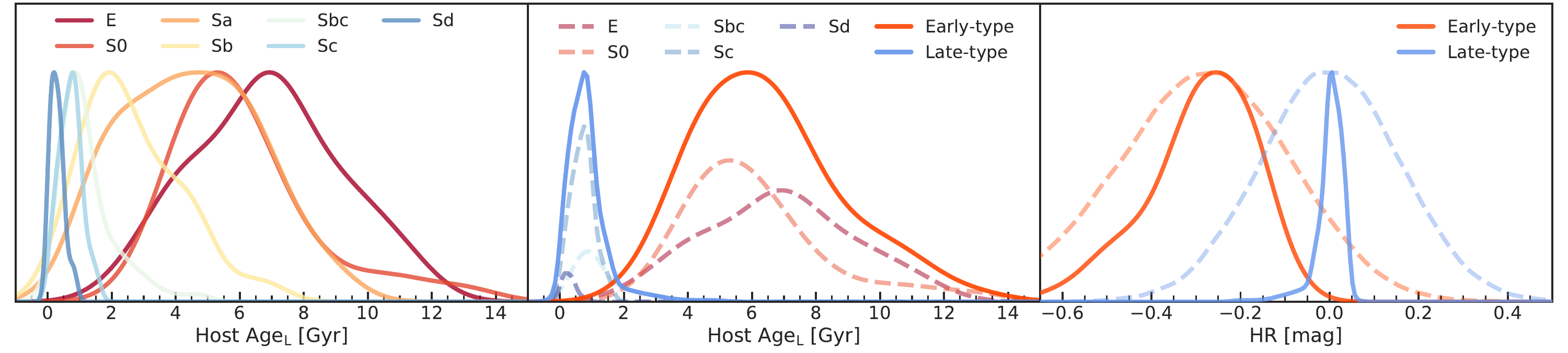}
        \caption{Morphology-binned distributions from CALIFA and our method to empirically determine HR distributions. Left: luminosity (flux)-weighted galaxy age for each Hubble type, normalised to equal height. Middle: age distributions binned for early and late types (solid), normalised to equal height. Contributions from original types, weighted by the number of host types in our SN~Ia sample, are shown with dashed lines. Right: distributions in HR space projected from host luminosity-weighted age using relations reported by \Kang. The HR-axis is offset so that the mean value of the late-type population is fixed at $\text{HR}=0$\,mag. Dashed lines in the right panel represent simulated Gaussian broadening due to intrinsic scatter in our SN~Ia sample, which is used in our analysis in Section~\ref{sec:results}.}
        \label{fig:distributions_process}
    \end{figure*}
    
    \subsection{Hubble Residuals from SNe~Ia}
    \label{sec:data-SNe}
    We source our SN~Ia dataset from the Second Amendment (A2) compilation \citep{Boruah_2020_A2} of 465 objects, which is itself comprised of SNe~Ia from numerous sources \citep{A1,G12,CSP3,Foundation}. In addition to convenience, our motivation for using A2 over directly sourcing observations from its constituent parts \citep[and perhaps adding omitted sources, e.g.,][]{S19} is that (i) A2 carefully checks (and updates where necessary) all host-galaxy redshifts, and, more importantly, (ii) it adjusts the distances of each subsample to bring the entire sample onto a common relative distance scale. The small redshifts of the SNe in our sample not only minimise the observational bias in sampling, but also ensure clear and reliable classifications of host-galaxy morphologies (see Sec.~\ref{sec:galaxy_crossmatch}).
    
    Although this sample is already viable for a cosmological analysis (having had undesirable objects filtered out), we further filter based on redshift (keeping only those 416/450 with redshift $z > 0.01$ to mitigate the effect of peculiar velocities) and distance-modulus uncertainty (requiring $\sigma_\mu < 0.25$\,mag; 413/416 objects satisfy this condition).
    We then fit a \rev{fiducial} \LCDM cosmology to the data to obtain Hubble residuals as $ \text{HR}_i = \mu_i - \mu_\text{cosmo}(z_i)$, though given the low redshifts involved the model has negligible dependence on cosmological parameters. These derived HR values will henceforth be our probe for SN~Ia luminosity. As is customary, the uncertainties ($\sigma_\text{HR}$) include an intrinsic scatter component that is determined simultaneously in fitting the cosmological model. 

    \setlength\tabcolsep{3pt} 
    \begin{table}
        \centering
        \begin{tabular}{|c|c||c|c|c|c|c|c|}
            \hline
            Source & Method & SN & Type & Mass & Age$_M$ & Age$_L$ & Morph \\
            \hhline{|=|=||=|=|=|=|=|=|}
            A2 & - & Y & - & - & - & - & LEDA\\
            \hline
            \Uddin & Z-PEG & Y &  P & Y & - & - & - \\
            \RoseMC & FSPS & Y & P & $(\alpha)$ & Y & - & - \\
            \Kang & YEPS & Y & S & - & - & Y & $(\alpha)$ \\
            \hline
            CALIFA & STARLIGHT & - & S & Y & Y & Y & direct\\
            SDSS DR8 & FSPS$^{(\beta)}$ & - & P & Y & Y & - & GZoo\\
            SDSS DR8 & Portsmouth & - & P & Y & Y & - & GZoo\\
            MaNGA & Pipe3D & - & S & Y & Y & Y & GZoo\\
            MaNGA & FIREFLY & - & S & Y & Y & Y & GZoo\\
            \hline
        \end{tabular}
        \footnotesize{$(\alpha)$: included but irrelevant to our analysis. $(\beta)$: `Granada' method.}
        \caption{A summary of the characteristics of datasets involved in this work. `Y' denotes the availability. `S' and `P' in the Type column denote Spectroscopic and Photometric data or fitting, respectively.}
        \label{tab:my_label}
    \end{table}
    
    \subsection{Host Galaxies and their Morphologies}
    \label{sec:galaxy_crossmatch}
    \rev{
    To determine the host galaxy of each SN~Ia in our sample, we query the Open Supernova Catalog \citep[OSC;][]{2017ApJ...835...64G}. Those without a host listed are then searched for in the NASA/IPAC Extragalactic Database (NED)\footnote{The NASA/IPAC Extragalactic Database (NED) is funded by the National Aeronautics and Space Administration (NASA) and operated by the California Institute of Technology.} using a region centred at their coordinates and extending in a $0.1' \times 0.1'$  square in right ascension and declination. From the resulting lists, we take the closest galaxy to be the host. If no matches are found within this window, we iteratively increase the search radius until a galaxy is located. The redshifts of these SNe~Ia are cross-referenced with those of the potential hosts, and those that agree are definitively determined to be hosts.
    }

    \rev{
    Of the 465 A2 objects (see Sec.~\ref{sec:data-SNe}), 427 hosts are listed in the OSC or in the original SN publications \citep[e.g.,][]{mlcs2k2,Hicken09, Ganeshalingam2010,CSP3,Foundation}. Five additional hosts are determined via proximity and cross-referenced redshifts. The remaining SNe~Ia either do not have listed hosts or were missing information (e.g., missing redshifts, potential host not found in any database). This results in host identifications for all objects except SN~2016dxv, whose host does not seem to be documented. Combined with the cuts described in Section~\ref{sec:data-SNe}, we obtain 373/465 objects with cross-matched host galaxies.
    }
    
    \rev{
    Following the acquisition of host-galaxy names, we obtain morphological types by cross-matching against HyperLEDA \citep{Makarov2014_HyperLEDA}, or if not available, from NED or SIMBAD \citep{2000A&AS..143....9W}.
    We limit the classification quality to the top two available groups (i.e., 0--1: ``detailed classification") to ensure reliability. 
    As a result, we retain 293 SNe~Ia with reliable host galaxies and corresponding morphological types. 
    }

    \subsection{SN Ia Luminosity models}
    \label{sec:luminosity_models}
    \rev{
    We test three SN Ia luminosity models taken from recent studies: \RoseMC (with \ZL), \Kang, and \Uddin.} 
    \rev{
    \RoseMC use FSPS \citep{Conroy_2009_FSPS} to estimate the mass-weighted age (age$_M$) from SED fitting to $ugriz$-band photomety. The slope between HR and age$_{M,\text{global}}$\footnote{In this study, our method only allows us to use the ``global" slope.} was reanalysed by \ZL and reported to be $-0.036\pm0.007$\,mag\,Gyr$^{-1}$.
    Similarly, but using luminosity-weighted age (age$_L$) spectroscopically determined with YEPS \citep{Chung_2013_YEPS}, \Kang reported an HR--age$_L$ slope of $-0.051\pm0.022$\,mag\,Gyr$^{-1}$.
    \Uddin studied the HR--mass relation using multiband photometric SED fitting with Z-PEG \citep{LeBorgne_2002_ZPEG}. Slopes are reported individually for distinct passbands, and so we take the mean of the reported values in optical and near-infrared (NIR) bands ($BgVri$)\footnote{As A2 is standardised differently than the method used by \Uddin, averaging across their (nearly consistent) optical and NIR results avoids bias.} for the same same redshift range as ours (i.e., $z>0.01$): $-0.0263\pm0.0257$\,mag\,dex$^{-1}$.
    }

    \begin{figure*}
        \centering
        \includegraphics[width=\linewidth]{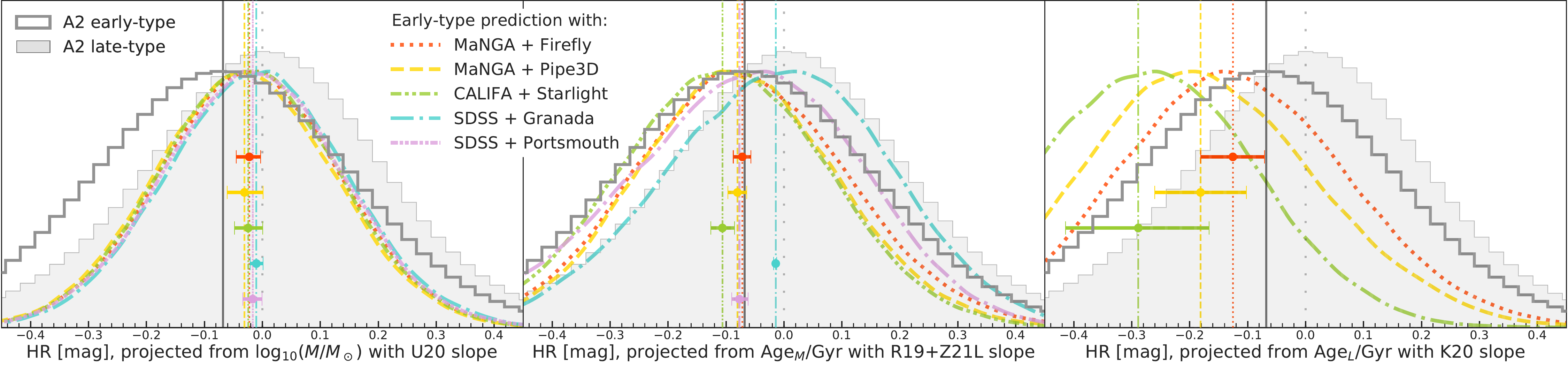}
        \caption{Left: SN~Ia distributions for early-type and late-type host morphologies (step histograms) and prediction of early-type population with \Uddin's HR--mass relation. Colours distinguish predictions from different data sets, and error bars show the location and the propagated uncertainty of late-type mean values. 
        HR offsets are applied so that the mean value for all late-type populations is zero. SN~Ia data are sampled with the size of the uncertainty, and the mean value from early-type hosts is shown with solid black. The agreement of all early-type means with this line confirms consistency and universality across data sets and EPS models. Middle: same as left, but with the HR--age$_M$ relation reported by \ZL. Right: same as left, but with the HR--age$_L$ relation reported by \Kang. 
        } 
        \label{fig:main_combined}
    \end{figure*}

    \subsection{Independent Galaxy Surveys}
    \rev{
    To empirically determine host properties for our morphologically binned sample, we use five value-added catalogues (VACs) for galaxy surveys that are completely independent from SN studies.
    The VACs provide the best-fit properties based on specific stellar population synthesis (SPS) models \citep[see, e.g.,][for a review]{Conroy_SPS_review}. Comparison between these five catalogues therefore allows us to test the effect of changing EPS models, data types (spectrostopic or photometric), and the source of data.
    }
    
    \rev{
    Our selected VACs are derived from three surveys with different sample sizes and characteristics: the Calar Alto Legacy Integral Field Area survey \citep[CALIFA;][]{CALIFA_2012_overview}, the Sloan Digital Sky Survey \citep[SDSS;][]{SDSS_III}, and SDSS's Mapping Nearby Galaxies at APO \citep[MaNGA;][]{MaNGA_overview}. 
    The CALIFA survey utilises integral field unit (IFU) spectroscopy.
    Its most recent data release \citep[DR3;][]{CALIFA_2016_DR3} and VAC \citep[PyCasso;][]{PyCasso_2017} provide galaxy properties for both spatially resolved local regions and global regions with integrated flux, analysed with the spectroscopic fitting code {\tt STARLIGHT} \citep{cidFernandes_STARLIGHT}. We use only the reported global properties.
    As a counterpart to the CALIFA VAC, we use the FIREFLY \citep{MaNGA_firefly} and Pipe3D \citep{MaNGA_pipe3d} VACs that analyse IFU spectra from MaNGA survey data. Similar to CALIFA, we use only the integrated-flux (``global") values.
    We further add two SED-fitting based VACs for SDSS's $ugriz$ photometry in DR8, provided by the Granada and Portsmouth groups \citep{SDSS_granada_portsmouth}. Owing to the nature of 5-band photometric SED fitting, these two VACs do not contain luminosity-weighted ages. Still, the large number ($\sim 0.2$ million) of samples provides statistically robust determinations of property distributions.
    }
    
    \rev{
    For all three datasets, the Hubble morphological types (or corresponding T-type values) are obtained externally and cross-matched. 
    For CALIFA galaxies, we match against morphological classifications from \cite{CALIFA_2015_morph}, who report direct and visual determinations. This results in our base dataset of 256 galaxies.
    Given the number of SDSS and MaNGA galaxies, direct visual classifications provided by professional astronomers are not available. Alternatively, Galaxy Zoo (GZoo) provides a successful classification of galaxies based on public votes. We use the linear model reported by \cite{GalaxyZoo2} for T-type estimation, based on the cross-matched GZoo dataset in both DR8 and MaNGA. 
    Since the final binning discussed in Section~\ref{sec:empirical_prediction} only distinguishes early and late types, the issue with the linear T-type model in GZoo having relatively large uncertainty should not significantly affect our results.  
    }

\section{Empirical Prediction of HR distribution}   
\label{sec:empirical_prediction}
    \rev{
    The morphology information in our A2 subsample and the independent galaxy-survey VACs allow us to simulate the distribution of host properties for our sample --- such as mass or age --- without directly observing them. Using those host-property distributions, we can predict the SN Ia luminosity (HR) distributions in different host-morphology bins via SN Ia luminosity models, such as the mass or age slopes in \Uddin, \RoseMC+ \ZL, and \Kang. 
    The resulting distributions are empirically predicted populations, reflecting even non-Gaussian features that are intrinsic to galaxies in the low-$z$ universe. Comparing such predictions with directly observed HR distributions therefore allows us to test the universality of proposed SN~Ia luminosity models.
    }
    
    \rev{
    The process of generating predicted distributions is shown in Figure~\ref{fig:distributions_process}.
    In each galaxy VAC, we use the cross-matched T-type or Hubble type as the bases of our empirically determined distributions in stellar masses, age$_{L}$, and age$_{M}$ (see, e.g., the left panel of Fig.~\ref{fig:distributions_process}, where we use kernel density estimation to represent the distribution).
    }
    
    \rev{
    We then add weight to the distribution for each Hubble type by the number of matching SNe in our sample, as shown in the middle panel in Figure~\ref{fig:distributions_process}. Since CALIFA does not include S0a, Sab, and Scd types, we regard them as the nearest types (in CALIFA labels). For instance, an Scd type in A2 is counted toward CALIFA's Sc type. S0a and Sab types are excluded in the final binning described below. The data are then binned into early-type (E and S0) and late-type (Sbc, Sc, and Sd) populations. The intermediate types (S0a, Sa, Sab, and Sb) are excluded to maximise the separation size.
    }
    
    \rev{Finally, the host-property distributions are projected to HR-space using the proposed/implied slopes in various studies. For example, in the right panel of Figure~\ref{fig:distributions_process}, we use the HR--age$_L$ relation from \Kang
    , $\Delta \text{HR} [\text{mag}] \approx -0.051 \cdot \Delta \text{Age} [\text{Gyr}]$. To incorporate the intrinsic scatter in this prediction, we simulate Gaussian broadening of the distribution based on the size of the intrinsic scatter obtained during the cosmology fitting shown in Figure~\ref{fig:HR_cosmo}. 
    The resulting curves are predicted HR-value distributions for SNe~Ia from early-type and late-type host galaxies, assuming the galaxy properties for each Hubble type in the utilised galaxy dataset are reliable.}

\section{Results}
\label{sec:results}
    \rev{
    The SN samples exhibit an HR separation --- the absolute difference between the samples' means --- of $0.068$\,mag between early- and late-type distributions. 
    These samples are calculated from 117 early- and 75 late-type sources, and sampled 10,000 times with uncertainties to build up the plotted distributions. 
    This separation has a small Welsch $t$-test\footnotemark{} $p$-value ($0.005$), and is therefore statistically significant.
    }
    
    \footnotetext{The Welsch $t$-statistic is calculated on each pair of the bootstrapped sample of the original 117 early-type and 75 late-type HR values keeping the sample size to the original data. The number of bootstrap samples is \num{50000}.}
    
    \rev{
    Given this significant separation, we compare the empirically predicted distributions against our SN~Ia data. These predicted populations are generated from independent galaxy surveys and projected onto the HR axis using specific SN Ia luminosity models as described in Section~\ref{sec:empirical_prediction}.
    Similarly to the evaluation of separation significance described above, we use the mean values of early- and late-type bins for each predicted distribution to evaluate the agreement with our A2 dataset. The results are shown in Figure~\ref{fig:main_combined}, 
    which omits predicted late-type distributions, since their mean values are fixed at zero similarly to the SN~Ia late-type sample.
    In this test, scatter in the predicted mean value in the early-type bin indicates the disagreement in galaxy properties between VACs, and the overall offset of predicted popuations from the A2 early-type mean indicates the underestimated or overestimated slopes in each SN~Ia luminosity model.
    }
    
    \rev{
    Figure~\ref{fig:main_combined} (left panel) shows the results with \Uddin's HR--mass slope. Most galaxy datasets are well in agreement, except the slight offset in ``Granada" (FSPS) VACs. However, the agreed predictions are low by a factor of $\sim3$, indicating an underestimated slope. 
    }
    
    \rev{
    Contrary to the test of HR--mass slope described above, the HR--age$_M$ slope test (middle panel in Fig.~\ref{fig:main_combined}), based on \RoseMC and \ZL, shows moderate agreement between the predicted early-type populations and our SN~Ia sample. While the CALIFA VAC's curve will also be brought into agreement if the slope is reduced by 15\%, the scatter among different galaxy datasets is more prominent than the mass estimations. In particular, the aforementioned Granada VAC's curve shows a significantly peculiar prediction, indicating a possible issue in the age$_M$ estimations. This issue is discussed in Section~\ref{sec: discussion}.
    }
    
    \rev{
    The result derived with the HR--age$_L$ relation by \Kang is shown in Figure~\ref{fig:main_combined} (right panel). The luminosity-weighted age is not provided\footnote{Owing to the known inaccuracy.} by both Granada and Portsmouth VACs; consequently, only three galaxy samples are used for the analysis. While the Firefly VAC's curve is at 1$\sigma$ to the SN sample's early-type mean, these three datasets indicate a possible overestimation of the slope by \Kang. Similarly to age$_M$, the predictions with different VACs are inconsistent.
    }
    
\section{Discussion: An Issue in the Current Debate}
\label{sec: discussion}
    Using \Nsamples low-redshift SNe~Ia, we have demonstrated that a small, yet nonzero separation in HR values exists between early-type and late-type bins of host-galaxy morphological types. 
    Our analysis also shows that, based on empirically determined distributions of galaxy properties, the nonzero slopes between SN~Ia luminosity and host-environment properties (e.g., HR--age, HR--mass) indeed predict such separations.
    \rev{
    Comparing those predicted separations to our SN data suggests that (i) \Uddin's optical-band HR--mass slopes are underestimated, (ii) \ZL's and therefore \RoseMC's slope is of the correct size, and (iii) \Kang's slope is overestimated.
    }
    
    \rev{Of the five galaxy VACs used in our analysis, the Granada catalogue for SDSS DR8 photometry exhibits peculiar behaviour, especially in age$_M$ space. This may be due to the known effect of the ``simple-tau" model as discussed by \RoseMC: the estimated age$_M$ may be significantly overestimated for young populations when a simple exponentially-decaying star-forming history (SFH) is employed. \RoseMC accounts for this by using a multiparametric SFH ``four-tau" model, but Granada samples are calculated with a simpler model. This explanation is consistent with the underestimated separation in the middle panel of Figure~\ref{fig:main_combined}, since the overestimated young populations can make the age offset between early-type and late-type galaxies significantly smaller. 
    }
    
    \rev{
    More fundamentally, the scatter in the predictions that is more prominent in both age$_L$ and age$_M$ spaces highlights the underlying issue in the current debate on SN~Ia luminosity and host properties. Host properties, such as mass or age, are \emph{very} model-dependent; using different SPS models can cause disagreement between predicted populations. We suggest that it is \emph{not} the effect of different galaxy samples that we are seeing as scatter in the results, since the predictions differ even for the same galaxy samples (e.g., Granada and Portsmouth).
    This is in fact a known issue, as stated by \cite{Chen_2010_EPS_comparison}. Estimated galaxy properties are strongly model-dependent, and therefore we cannot expect them to be an unbiased portrayal of physical truth. \cite{Longhetti_2009_EPSmass} and \cite{Lee_2007_EPSage} show that the spectral synthesis of galaxies is not as sensitive to age as it is to mass, suggesting that age estimation is expected to be much less accurate and inconsistent.
    The discussion in the SN~Ia literature on this issue has been insufficient, since the effect is not significant enough to cause severe disagreements in mass-space, as seen in our results as well.
    Therefore, all proposed models are only applicable and relevant to studies that employ the \emph{exact} methods for property estimations of galaxies, and comparison between differently derived models, such as \Kang and \RoseMC (as done by \Lee), is not only problematic due to the mixed properties (age$_M$ and age$_L$) but also in terms of different SPS models that are employed in these studies.
    }
    
    \rev{
    Our results call into question certain details of the now accepted mass-step: though less apparent and significant, mass estimates can vary if different SPS models are used, often in a correctable way (e.g., constant offsets or linear relation).
    This suggests that one can reevaluate the previous (dis)agreement on mass-step location or size, ultimately propagating the newly understood bias or uncertainty to the cosmological analysis to which the mass-step is applied.
    }
    
    \rev{
    SN~Ia and host-galaxy property correlations do indeed exist, but the systematic bias introduced by SPS models needs to be understood before applying such correlations to cosmological analyses.
    We will discuss and quantify this SPS-model systematics in our next work (Murakami et al., in prep.).
    }

\section*{Acknowledgements}
    \rev{We thank our referee, Benjamin Rose, whose feedback led to a substantial improvement of our initial manuscript. Y.S.M. is also grateful to Mariska Kriek for an eye-opening conversation about SPS models and the diversity in predictions they produce.}
    We thank Thomas de Jaeger for helpful discussions on the mass-step and Andrew Hoffman for proofreading.
    A.V.F.'s group at U.C. Berkeley acknowledges generous support from Marc J. Staley, the Christopher R. Redlich Fund, Sunil Nagaraj, Landon Noll, the TABASGO Foundation, and the Miller Institute for Basic Research in Science (U.C. Berkeley).
    We acknowledge the use of data from SDSS and CALIFA: the full acknowledgement texts are available at our GitHub link below.

\section*{Data Availability}
    The raw data used in our analysis are available from cited papers. The tools we used in our analysis and the links to VACs are available at \href{https://github.com/SterlingYM/SN_Ia_and_Host_properties}{https://github.com/SterlingYM/SN\_Ia\_and\_Host\_properties}.


\bibliographystyle{mnras}
\bibliography{main.bib}



\label{lastpage}
\end{document}